\pdfoutput=1
\documentclass[sigconf]{acmart}

\renewcommand\footnotetextcopyrightpermission[1]{} 
\setcopyright{none}                 
\acmConference[SIGIR '26]{Proceedings of the 49th International ACM SIGIR Conference on Research and Development in Information Retrieval}{July 20--24, 2026}{Melbourne, VIC, Australia}
\acmBooktitle{Proceedings of the 49th International ACM SIGIR Conference on Research and Development in Information Retrieval (SIGIR '26), July 20--24, 2026, Melbourne, VIC, Australia}

\usepackage{graphicx}

\usepackage{enumitem}
\usepackage{balance} 
\usepackage{hyperref}
\begin{document}

\title{Beyond Item IDs: Scaling Short-Form-Video Recommendation via Semantic-Native Long Sequence Modeling}



\author{Ruixiao Sun}
\affiliation{
  \institution{Google}
  \country{Mountain View, USA}
}

\author{Diego Uribe Mora}
\affiliation{
  \institution{Google}
  \country{Mountain View, USA}
}
\author{Zhimeng Jiang}
\affiliation{
  \institution{Google}
  \country{Mountain View, USA}
}
\author{Yuanzhen Lin}
\affiliation{
  \institution{Google}
  \country{Mountain View, USA}
}
\author{Jiarui Wang}
\affiliation{
  \institution{Google}
  \country{Mountain View, USA}
}
\author{Yuening Li}
\affiliation{
  \institution{Google}
  \country{Mountain View, USA}
}
\author{Danfeng Guo}
\affiliation{
  \institution{Google}
  \country{Mountain View, USA}
}
\author{Zhizhong Chen}
\affiliation{
  \institution{Google}
  \country{Mountain View, USA}
}
\author{Chuan He}
\affiliation{
  \institution{Google}
  \country{Mountain View, USA}
}
\author{Liang Liu}
\affiliation{
  \institution{Google}
  \country{Mountain View, USA}
}

\renewcommand{\shortauthors}{Ruixiao Sun et al.}

\begin{abstract}
Capturing user interests across extensive watch histories is critical for short-form video recommendation, yet scaling sequence length is limited by two bottlenecks: the semantic sparsity of atomic Video IDs and the quadratic computational complexity of Transformers. Traditional orthogonal Video IDs fail to capture content relationships and demand large embedding tables, while the quadratic complexity of self-attention restricts the maximum sequence length under strict industrial latency and resource constraints. In this work, we present a production-deployed framework for modeling ultra-long user behavior sequences at a billion-user scale.
We first address the representation bottleneck by adopting content-native Semantic IDs. By utilizing depth-truncated, coarse-grained Semantic IDs, we shrink the embedding table size from corpus cardinality. This compact representation naturally generalizes to cold-start content through shared semantic prefixes. Second, to overcome the sequence scaling barrier, we introduce a Global-Aware Compression Transformer that leverages non-parametric temporal folding and unified global query integration to effectively condense the sequence, alleviating both the memory and computational bottlenecks of standard self-attention. Offline profiling on our computing infrastructure demonstrates an order-of-magnitude reduction in peak memory footprint and a drastic decrease in computational overhead. This efficiency gain enables supporting longer sequence lengths at an affordable cost in production, yielding substantial online gains in satisfied user engagement and satisfied content consumption in large-scale online A/B tests.
\end{abstract}

\ccsdesc[500]{Information systems~Recommender systems}

\ccsdesc[500]{Information systems~Personalization}

\keywords{Semantic Representation, User Long Sequence Modeling, Efficient Transformer, Large Scale Recommendation System}
\maketitle


\section{Introduction}
Sequential user behavior modeling is a cornerstone of modern recommendation systems, particularly in short-form video feeds where sequences of length $L > 10^3$ are necessary to encapsulate both long-term stable preferences and immediate short-term intent \cite{Covington2016Youtube, Hidasi2016Session}. However, scaling to these lengths presents a dual challenge: a representation bottleneck caused by traditional item indexing and a computational bottleneck inherent in Transformer architectures.

Sequential recommendation has evolved from early RNNs \cite{Hidasi2016Session, donkers2017sequential} to self-attention mechanisms \cite{kang2018self, sun2019bert4rec}, but the quadratic complexity of Transformers necessitated two-stage industrial solutions. Beyond target-aware attention (DIN \cite{zhou2018deep}, DIEN \cite{zhou2019deep}), methods scaling to longer histories employ either heuristic filtering (SIM \cite{pi2020search}, TWIN \cite{chang2023twin}) or hierarchical clustering (TWIN-V2 \cite{si2024twinv2}). However, these decoupled strategies either incur information loss by discarding signals or require complex offline pre-computation.

Consequently, recent efforts have pivoted toward optimizing architectures to model full sequences directly in an end-to-end manner. Following the early validation of Transformers in BST \cite{chen2019bst}, recent industrial powerhouses have introduced sophisticated optimizations to push the sequence length boundaries. For instance, ByteDance introduced LONGER \cite{longer2025} to scale sequence length and HyFormer \cite{hyformer2026} to unify sequential behaviors with non-sequential interactions. Simultaneously, Meta pushed the boundaries of generative and representation learning with DV365 \cite{lyu2025dv365}, HSTU \cite{hstu2024} and VISTA \cite{chen2025massive} for universal user sequence representation.

\textbf{The Semantic-Native Representation Gap.} Parallel to architectural scaling, item representation is undergoing a paradigm shift. Early solutions like TDM \cite{zhu2018tdm} and Deep Retrieval \cite{gao2020deepretrieval} relied on hashed IDs, presenting two major flaws: (1) Semantic fragmentation: Treating IDs as orthogonal tokens forces models to "memorize" interactions, leading to poor generalization on cold-start content. (2) Storage and sparsity: Hashed IDs create prohibitively large embedding tables that scale unboundedly with corpus size. Recently, the field has evolved towards content-native Semantic Identifiers (SIDs) via RQ-VAE \cite{rajput2023recommender}. While works like PLUM \cite{plum2025} and TRM \cite{zhao2026farewell} explore hierarchical SIDs, their synergy with ultra-long sequence modeling remains under-explored.

In this work, we present a framework that bridges this gap by synergizing the generalization power of hierarchical SIDs with a Global-Aware Compressed Transformer. By adopting content-native SIDs, we enhance the model's generalization from semantic representation. Concurrently, we leverage a parameter-free temporal folding strategy that explicitly trades temporal resolution for feature dimensionality. This structural reorganization allows us to scale user histories to thousands of interactions while drastically reducing memory footprint, effectively supporting production deployment. To summarize, our main contributions are as follows: 
\begin{itemize}[leftmargin=*]
    \item \textbf{Semantic-Native Representation at Scale:} We present a large-scale industrial deployment of SIDs based on RQ-VAE  within a billion-user recommendation system. By implementing a Depth-Truncated Bi-gram strategy, we effectively resolve the vocabulary and sparsity bottlenecks inherent in traditional Video IDs. This approach decouples embedding storage from corpus size while significantly improving generalization to cold-start content.

    \item \textbf{High-Expressivity Efficient Sequence Architecture:} We design a Global-Aware Compressed Transformer that leverages non-parametric temporal folding to condense sequence length. This achieves a performance-preserving reduction in computational complexity while simultaneously enhancing representational expressivity. By stacking windows into super-tokens, the model expands the feature dimension to capture complex local interactions within ultra-long user sequences.

    \item \textbf{Large-Scale Industrial Validation:} We provide rigorous offline and online evaluations on a major short-video platform serving billions of users. Our results demonstrate a substantial reduction in both peak memory and computational cost within the sequence modeling component. These efficiency gains allow us to scale sequence lengths, yielding significant online gains in satisfied watch time.
\end{itemize}
\section{Methodology}
We present a high-performance sequence modeling framework deployed on a global short-video platform serving billions of active users. The system is designed to capture user interests over horizons spanning thousands of interactions while adhering to strict millisecond-level inference latency constraints.

\subsection{System Architecture}
To reconcile real-time responsiveness with long-term memory, our system adopts Hybrid Synchronous-Asynchronous Framework\cite{li2024short}, decoupling user interests modeling into two modules:
\begin{itemize}[leftmargin=*]
    \item \textbf{Synchronous Short-Term Module:} A real-time component embedded in the serving graph. It encodes the immediate interaction tail $\mathcal{S}_{short}$ to capture emerging interests via real-time feature ingestion.
    \item \textbf{Asynchronous Long-Term Module:} An asynchronous component modeling the ultra-long user sequence $\mathcal{S}_{long} = \{v_1, \dots, v_L\}$. Crucially, by utilizing a shared backbone and optimization objective in our core ranking model system, we ensure that asynchronously computed representations are intrinsically mapped to a shared latent space. This prevents feature misalignment between long-term historical signals and real-time ranking features.
\end{itemize}

\subsection{Semantic-Native Representation}
To transcend the sparsity and cold-start limitations of traditional orthogonal Video IDs, we transition to a content-native paradigm utilizing the Semantic Identifier (SID) infrastructure \cite{rajput2023recommender}.

\subsubsection{Infrastructure: Hierarchical Quantization}
Our system ingests SIDs generated via an upstream Residual Quantized VAE (RQ-VAE) framework. This infrastructure projects high-dimensional, multi-modal content embeddings into a discrete, hierarchical latent space. Specifically, the RQ-VAE produces a tuple of discrete codes $s_v = (c_1, c_2, \dots, c_D)$ for each video, encoding a coarse-to-fine semantic hierarchy where initial codes capture broad categories and subsequent codes encode granular residuals.

\subsubsection{Strategic Adaptation: Asymmetric Semantic Modeling}

Using full-depth tokens for ultra-long user sequences ($\mathcal{S}_u$) is computationally prohibitive due to vocabulary explosion. We therefore implement an Asymmetric Granularity Strategy:

\begin{itemize}[leftmargin=*]
    \item \textbf{Target \& Non-sequence Side (Fine-Grained):} Both Candidate items and pivotal interaction features (e.g., the last-watched ID) retain extended semantic hierarchies for precise discriminative matching.
    \item \textbf{Sequence Side (Coarse-Grained):} Videos in user sequence are restricted to a Depth-Truncated Bigram $(c_1, c_2)$.
\end{itemize}

\begin{figure}[htbp]
  \centering
  \vspace{-2mm}

  \includegraphics[width=0.9\linewidth]{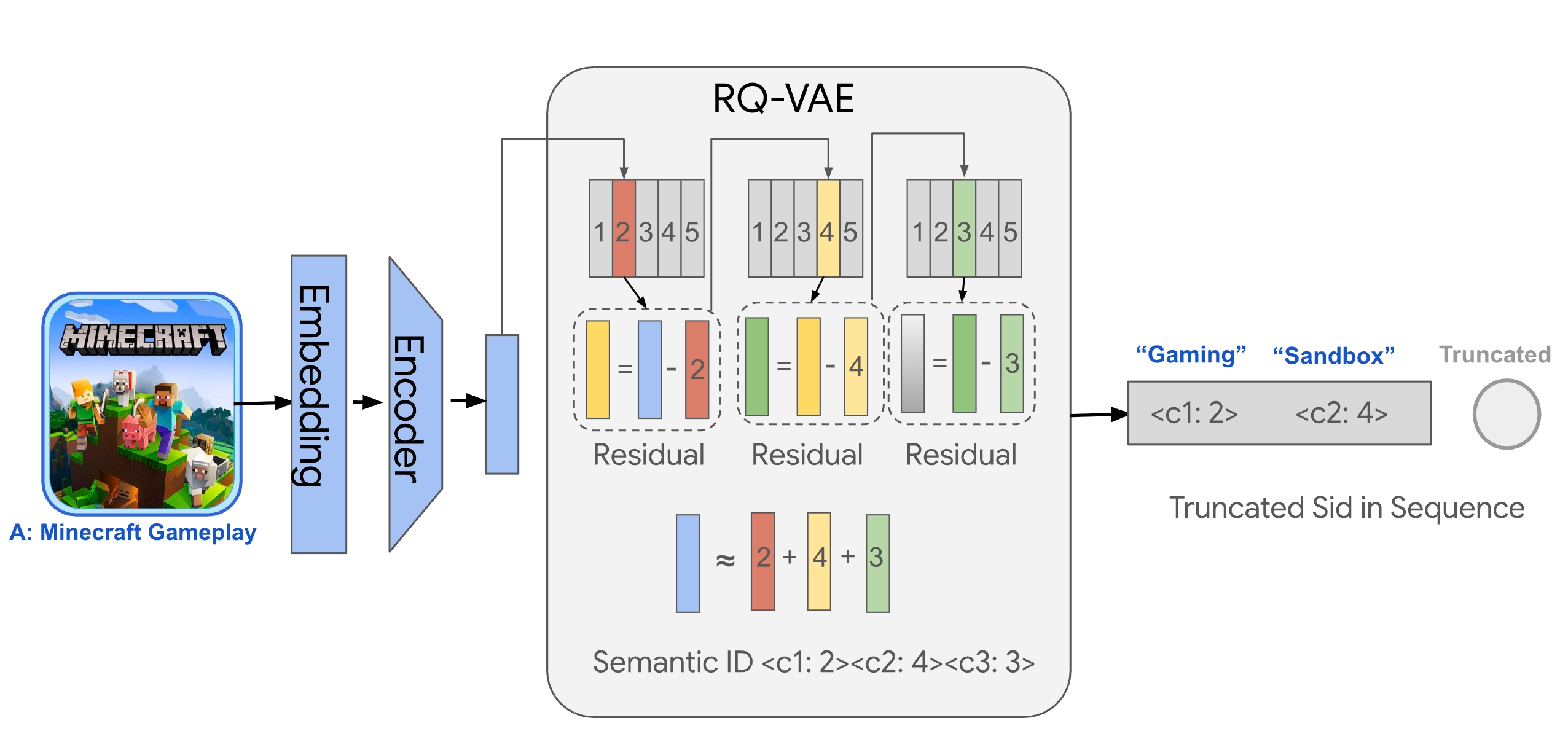}
  \caption{Semantic ID generation. Raw videos are quantized via RQ-VAE, followed by Bi-gram Truncation to flatten the path into a unified integer ID, preserving category-level semantics for sequence modeling.}
  \label{fig:sid_flow}
  \vspace{-3mm}
\end{figure}

As illustrated in Figure \ref{fig:sid_flow}, we project the vertical ($c_1$) and sub-vertical ($c_2$) codes into a unified identifier space 
$ID_{seq} = c_1 \cdot |V| + c_2,$ where $|V|$ denotes the codebook vocabulary size.

This design achieves a critical efficiency-expressivity trade-off. Unlike atomic Video IDs that scale unboundedly with corpus expansion, this approach bounds the embedding table size by regulating hierarchy depth. It drastically reduces memory footprint while preserving the semantic gist, effectively clustering interactions into discriminative sub-verticals (e.g., "Gaming-Sandbox" vs. "Gaming-Shooter") without the burden of full-depth modeling.

\subsection{Global-Aware Compressed Transformer}
\begin{figure}[htbp]
  \centering
  \vspace{-1mm}
  \includegraphics[width=0.9\linewidth]{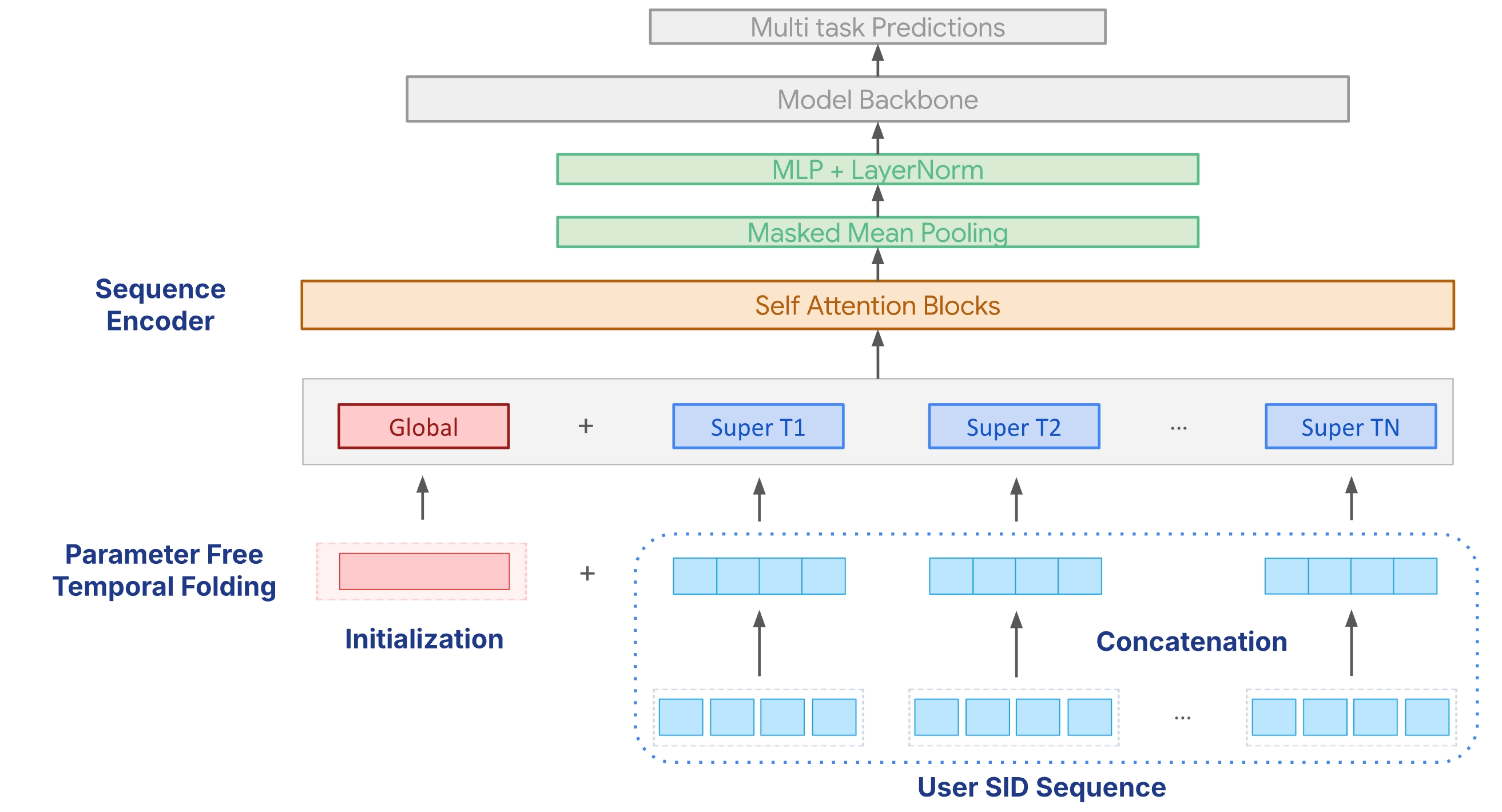}
  \caption{Global-Aware Compressed Transformer Overview.}
  \vspace{-4mm}
  \label{fig:model_arch}
\end{figure}

Standard self-attention incurs $O(L^2)$ complexity, which is prohibitive for $L \sim 10^3$. We propose a \textbf{Global-Aware Compressed Transformer}, as shown in Figure \ref{fig:model_arch}, that ensures computational tractability through lossless structural reorganization.

\subsubsection{Parameter-Free Temporal Folding}
Unlike pooling methods that reduce sequence length via lossy aggregation, Temporal Folding performs a non-destructive reshaping to mitigate quadratic complexity. We partition the input $\mathbf{X} \in \mathbb{R}^{L \times d}$ into windows of size $k$ and stack them channel-wise to form super-tokens:
\begin{equation}
    \mathbf{H}_{i}^{super} = \text{Concat}(\mathbf{x}_{(i-1)k+1}, \dots, \mathbf{x}_{ik})
\end{equation}

This zero-parameter operation transforms the input from $(L, d)$ to $(L/k, k \cdot d)$, explicitly trading temporal resolution for feature expressivity to capture dense dependencies between "super-events".

\subsubsection{Global-Local Representation Fusion}
To synthesize a holistic user representation, we prepend a learnable Global Query Token $\mathbf{G}_{init} \in \mathbb{R}^{N_g \times d}$ to the compressed sequence input. Distinguishing our approach from user profile-based initialization, we instantiate $\mathbf{g}_{init}$ as a strictly feature-agnostic anchor. This forces the attention mechanism to derive representations solely from dynamic interactions. Furthermore, acting as an attention sink \cite{xiao2023streaming}, it naturally absorbs low-entropy signals, stabilizing optimization.
We fuse this global state ($\mathbf{g}_{final}$) with granular sequential signals via Unified Masked Mean Pooling. Let $\mathbf{H}_{out} \in \mathbb{R}^{(L' + N_g) \times d}$ denote the complete Transformer output, where index $0$ corresponds to $\mathbf{g}_{final}$. The final representation is computed as:
\begin{equation}
    \mathbf{u}_{final} = \frac{\sum_{j=1}^{L' + N_g} (\mathbf{H}_{out}[j] \cdot \mathcal{M}'_j)}{\sum_{j=1}^{L' + N_g} \mathcal{M}'_j + \epsilon}
\end{equation}
where $\mathcal{M}' \in \{0, 1\}^{L' + N_g}$ is the extended validity mask ($\mathcal{M}'_{[1:N_g]} = 1$). This operation effectively balances global intent with local details, while the normalization ensures magnitude invariance across diverse history lengths.

\section{Experiments}
\subsection{Experimental Setup}
We evaluate our framework on the production traffic of a major short-form video platform. Our experiments are designed to validate the effectiveness of the Semantic IDs, the scalability of the sequence modeling architecture, and the system efficiency of our compression strategy. Finally, we measure the end-to-end business impact through large-scale online A/B experiments.

\subsection{Semantic-Native Representations Eval}
We first investigate the impact of replacing discrete Video IDs with Bi-gram SIDs in sequence modeling. The primary goal is to verify that content-native representations can match the memorization capability of ID-based baselines while improving the generalization of recommendation system.

\textbf{Online Ablation Results.}
Table \ref{tab:rep_validity} presents the relative lift against the Video ID baseline with sequence length $L=160$. 
The transition to SIDs achieves statistically significant positive lift on core engagement metrics (satisfied views). This confirms that compressed SID preserve the specific interaction signals previously captured by individual Video IDs.
Crucially, SIDs drive $+6.81\%$ gain in Freshness (satisfied views with recently uploaded content). This confirms that content-native representations effectively alleviate the cold-start problem, improving generalization of recommendation system. Furthermore, by bounding the vocabulary to the Bi-gram space, we achieve a substantial reduction in both embedding parameters and physical storage size

\begin{table}[h]
\centering
\vspace{-2mm}
\caption{Online Ablation Test ($L=160$). Semantic IDs yield positive top-line metrics and significant cold-start gains (all $p < 0.05$).}
\label{tab:rep_validity}
\begin{tabular}{l|c}
\toprule
\textbf{Metric} & \textbf{Relative Lift ($\Delta\%$)} \\
\midrule
Satisfied Views      & \textbf{+0.86\%} \\
Freshness         & \textbf{+6.81\%} \\
Embedding Parameters   & \textbf{-38.07\%} \\
Storage Size   & \textbf{-39.33\%} \\
\bottomrule
\end{tabular}
\vspace{-4mm}
\end{table}

\subsection{Scalability Analysis}
With the representation bottleneck addressed, we next validate how the Global-Aware Compressed Transformer handles ultra-long sequences. We first assess whether the model capacity is sufficient to capture complex user interests, and then investigate the gains from scaling the input sequence length.

\textbf{Impact of Model Capacity.}
We examine our model's scaling capacity by conducting experiments on the model dimensions (Hidden Dimension $d$ and number of transformer layers $N$). 
Table \ref{tab:capacity_scaling} demonstrates that the sequence modeling with SID benefits significantly from deeper and wider networks. The consistent AUC lift confirms that the architecture effectively utilizes additional parameters to capture intricate interaction patterns, validating its capability to scale with available compute budget.

\begin{table}[h]
\centering
\vspace{-2mm}
\caption{Model Capacity Scaling ($L=800$). The architecture benefits consistently from expanded width and depth across tasks.}
\label{tab:capacity_scaling}
\begin{tabular}{cc|cc}
\toprule
\textbf{Hidden Dim ($d$)} & \textbf{Layers ($N$)} & \textbf{Task 1 $\Delta$ AUC} & \textbf{Task 2 $\Delta$ AUC} \\
\midrule
\multicolumn{4}{c}{\textit{Impact of Hidden Dimension (Fixed Layers = 2)}} \\
\midrule
128 (Base) & 2 & -- & -- \\
256 & 2 & +0.04\% & +0.00\% \\
1024 & 2 & +0.10\% & +0.15\% \\
\midrule
\multicolumn{4}{c}{\textit{Impact of Layer Depth (Fixed Dim = 128)}} \\
\midrule
128 (Base) & 2 & -- & -- \\
128 & 4 & +0.16\% & +0.27\% \\
128 & 6 & +0.23\% & +0.36\% \\
\bottomrule
\end{tabular}
\vspace{-2mm}
\end{table}

\begin{figure}[ht]
    \centering
    \vspace{-1mm}
    \includegraphics[width=0.4\textwidth]{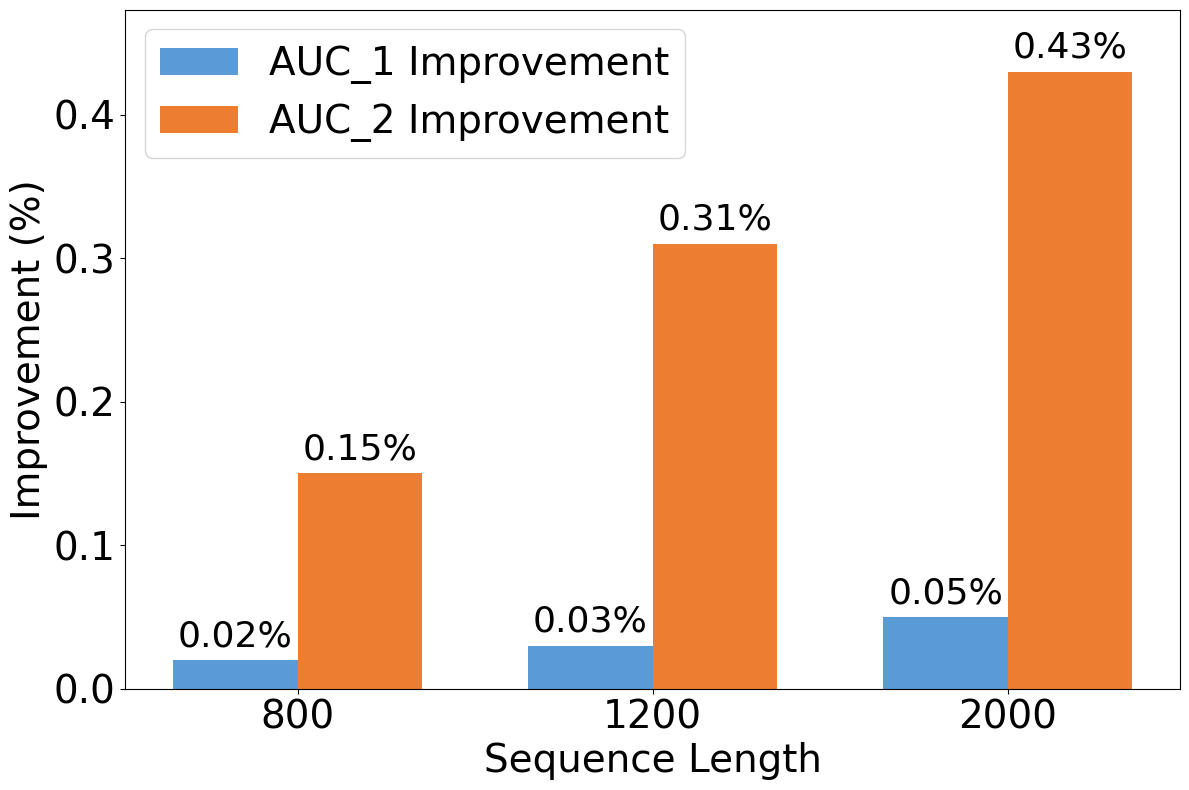}
    \caption{AUC Improvements vs. Sequence Length.}
    \label{fig:auc_longer}
    \vspace{-2mm}
\end{figure}

\begin{figure}[ht]
    \centering
    \vspace{-3mm}
    \includegraphics[width=0.45\textwidth]{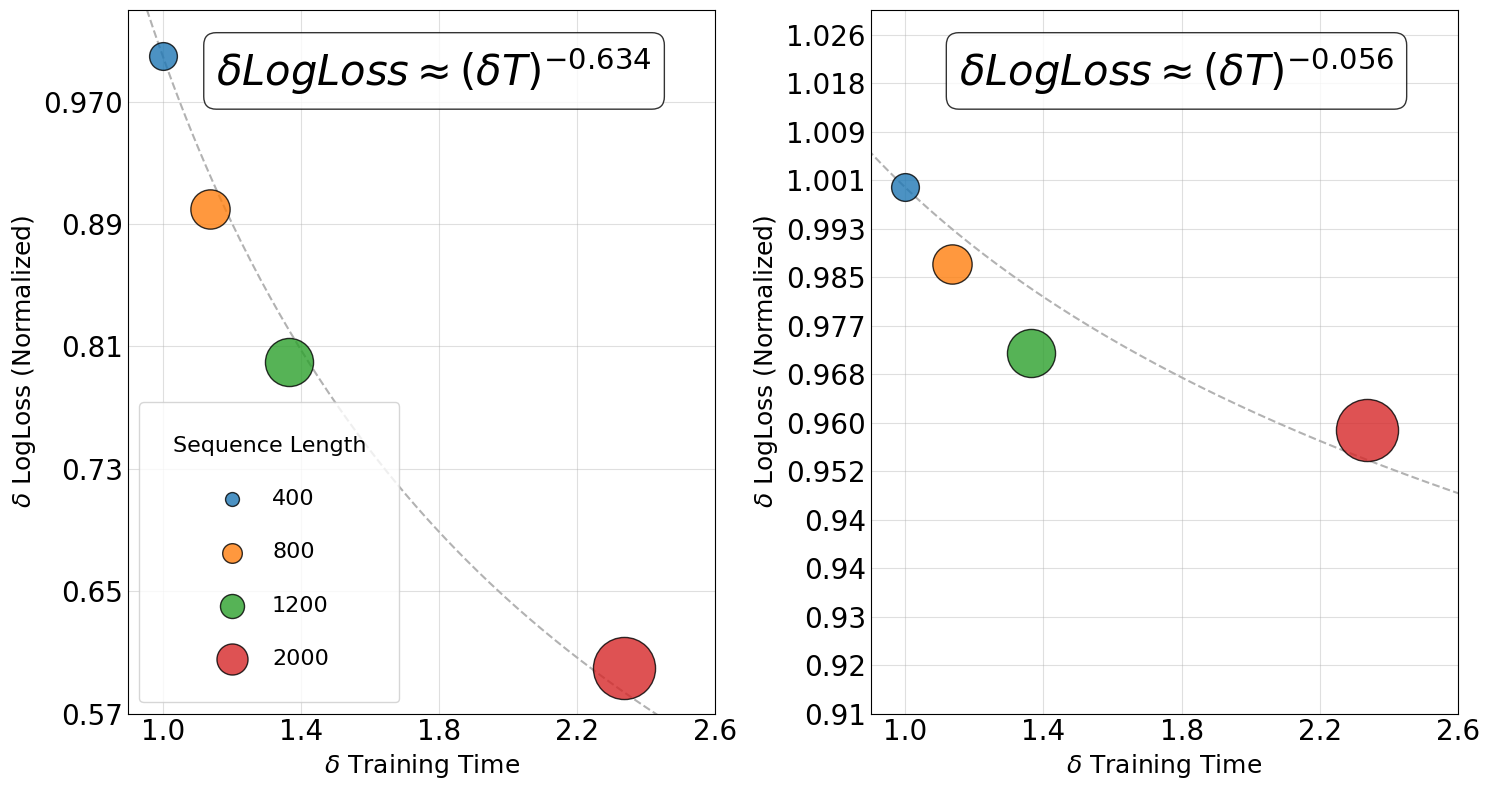}
    \caption{Empirical Normalized Scaling Law Analysis: The power-law fit (dashed lines) follows $\delta LogLoss \propto (\delta T)^k$, with scaling exponent $k$, relative loss $\delta LogLoss= \frac{LogLoss_{L}}{LogLoss_{400}}$, $\delta T= \frac{T_{L}}{T_{400}}$.}
    \label{fig:scaling_law_analysis}
    \vspace{-4mm}
\end{figure}

\textbf{Impact of Sequence Length.}
We now examine how the sequence model with SID performance scales with the user watch history length in \{400, 800, 1200, 2000\}. As shown in Figure \ref{fig:auc_longer}, we observe a consistent monotonic increase in offline AUC for different tasks. The gain from $L=1200$ to $L=2000$ indicates that ultra-long historical interactions contain valuable signals that are not captured by shorter windows with training speed drawdown. Figure \ref{fig:scaling_law_analysis} also shows power-law scaling function of relative LogLoss w.r.t. training time in the form $\frac{LogLoss_{L}}{LogLoss_{400}}=\frac{(T_{L}/D)^k}{(T_{400}/D)^k}=(\frac{T_{L}}{T_{400}})^k$. The exponential coefficients for task 1 and 2 are -0.634 and -0.056, respectively.

\subsection{Efficiency \& Performance Analysis}
To validate the trade-off between computational cost and model accuracy, we profile the sequence modeling component ($L=800$) on 64 chips within our computing infrastructure.

As shown in Table \ref{tab:efficiency}, our Global-Aware Compression ($k=4$) achieves a Pareto improvement. Specifically within the sequence modeling module, it reduces the training step time by 83.9\% and peak activation memory by 92.2\%, while lifting overall AUC by +0.06\% (Task 1) and +0.15\% (Task 2). We attribute this accuracy gain to local denoising and expanded feature expressivity resulting from temporal folding. By directly concatenating $k$ neighboring tokens (where $k$ is a tunable hyperparameter), the effective hidden dimension increases by a factor of $k$, allowing the attention mechanism to capture richer local feature correlations.

However, further increasing compression to $k=6$ causes an AUC regression (Task 1 $\Delta$: -0.04\%). This confirms that $k=4$ represents the optimal \textbf{efficiency-resolution trade-off}—sufficiently aggressive to unblock computational bottlenecks, yet retaining the temporal resolution needed to distinguish fine-grained sequential patterns.

\begin{table}[h]
\centering
\vspace{-2mm}
\caption{Efficiency \& Cost Profile ($L=800$). Temporal Folding ($k=4$) drastically reduces computational cost of the sequence component while slightly improving AUC on multiple tasks.}
\label{tab:efficiency}
\begin{tabular}{l|c|c}
\toprule
\textbf{Metric} & \textbf{Vanilla Transformer} & \textbf{Ours ($k=4$)} \\
\midrule
Task 1 $\Delta$ AUC & -- (Base) & \textbf{+0.06\%} \\
Task 2 $\Delta$ AUC & -- (Base) & \textbf{+0.15\%} \\
\midrule
Training Step Time (ms) & 41.1 & \textbf{6.6 (-83.9\%)} \\
Peak HBM (MiB) & 5758 & \textbf{448 (-92.2\%)} \\
\bottomrule
\end{tabular}
\vspace{-4mm}
\end{table}

\subsection{Full System Production Impact}
Finally, we presented the online A/B experiment results utilizing Semantic IDs and Global-Aware Compression to model ultra-long sequences ($L=2000$)—against the previous state-of-the-art production baseline ($L=800$ Video ID).

Notably, the superior efficiency of SIDs and our compression strategy released significant computational headroom. This allowed us to scale up the model capacity within the asynchronous modeling component, further boosting the system's ability to capture complex user interests.
Table \ref{tab:final_online} reports top-line metrics lift by this holistic system upgrade. 
By synergizing content-native representations with a deeper, longer-context architecture, the new system yields substantial gains across all key metrics. 
The significant lift in Actively Engaged Users, Satisfied Watch Time and Satisfied Views confirms that the upgraded system captures richer, long-term user interest signals that were previously inaccessible to the baseline model.

\begin{table}[h]
\centering
\vspace{-2mm}
\caption{Full System Production Impact. Comparison between the baseline ($L=800$ Video ID) and the fully optimized system ($L=2000$ SID). All reported cumulative gains from the system upgrade are statistically significant ($p<0.05$).}
\label{tab:final_online}
\begin{tabular}{l|c}
\toprule
\textbf{Metric} & \textbf{Total Relative Lift ($\Delta\%$)} \\
\midrule
Actively Engaged Users & \textbf{+0.52\%} \\
Satisfied Watch Time       & \textbf{+1.42\%} \\
Satisfied Views            & \textbf{+1.08\%} \\
\bottomrule
\end{tabular}
\vspace{-4mm}
\end{table}

\section{Conclusion}
In this work, we presented a scalable framework for modeling ultra-long user sequences using Semantic IDs, effectively resolving vocabulary bottlenecks and improving system generalization. To address computational constraints, we introduced a Global-Aware Compression-based Transformer, which leverages non-parametric temporal folding to significantly reduce memory footprint. This synergy enabled us to scale model capacity within a billion-user short video recommendation system, resulting in substantial cumulative gains in top-line metrics such as satisfied watch time and satisfied views. Our approach provides a practical, efficient blueprint for next-generation infinite-context recommendation engines. Moving forward, we aim to push the boundaries of efficient long-context kernels and hardware-algorithm co-design, further investigating sequence scaling laws as we transition into the era of Large Recommendation Models (LRMs).

\section*{Main Presenter Bio}
Dr. Ruixiao Sun is a Tech Lead and Machine Learning Scientist at Google, specializing in recommendation systems. Her work primarily focuses on next-generation recommendation systems, including Large Recommendation Models, multi-modal recommendations, LLM-inspired architectures, and user sequence modeling. 

\noindent\textbf{Presentation Preference:} Talk.

\bibliographystyle{ACM-Reference-Format}
\balance
\bibliography{main}

\end{document}